\documentclass[twocolumn,prl,superscriptaddress,preprintnumbers,floatfix,aps]{modirevtex4}
\usepackage{epsfig}
\usepackage{graphicx}
\usepackage{amsfonts,amssymb,amsmath}

\begin{document}

\title{First-Principles Studies of the Metallization and the Equation of State of Solid Helium}
\author{S.~A.~Khairallah$^1$ and B.~Militzer$^{1,2}$}
\affiliation{Departments of Earth and Planetary Science$^1$ and Astronomy$^2$, University of California, Berkeley, CA 94720, USA}

\begin{abstract}
The insulator-to-metal transition in solid helium at high pressure is
studied with first-principles simulations. Diffusion quantum Monte
Carlo (DMC) calculations predict that the band gap closes at a density
of $21.3$ g/cm$^{3}$ and a pressure of $25.7$ terapascals, which is
$20$\% higher in density and $40\%$ higher in pressure than predicted
by density functional calculations based on the generalized gradient
approximation (GGA). The metallization density derived from GW
calculations is found to be in very close agreement with DMC
predictions. The zero point motion of the nuclei had no effect on the
metallization density within the accuracy of the calculation. Finally,
fit functions for the equation of state are presented and the
magnitude of the electronic correlation effects left out of the GGA
approximation are discussed.
\end{abstract}

\maketitle

At low pressure, helium is an inert gas that exhibits a very large
electronic excitation gap of $19.8$ eV and has the highest
ionization energy of all atoms, $24.6$ eV. This is because helium has no
core electrons, so its valence electrons are bound more strongly than
in heavier atoms where screening effects play a role. Given such
a strong binding, extreme pressures are needed to reach
metallization. In fact, after neon~\cite{boettger,needs06}, solid helium is
expected to have the highest metallization pressure among all
elemental solids.

Metallic solid helium is expected to be present in the outer layers of
white dwarfs (WD)~\cite{chab05}. After the initial star has exhausted
all its nuclear fuel, it sheds its outer layer and leaves behind a
dense carbon-oxygen core of the size of the earth that is surrounded
by an envelope of pure helium, hydrogen, or a mixture. The fossil star
then spends the remaining of its lifetime cooling until vanishing
luminosity. Measuring the luminosity of the oldest WD would therefore
constrain the age of the galaxy~\cite{Oswalt}, which qualifies WDs as
stellar chronometers. 

Extracting the correct physics from WDs depends on how consistent the
cooling models are with the observed luminosity~\cite{isern}.
Characterizing helium at high pressure is important because its
properties regulate the heat transport across the outer layers. The
metallization transition is important because it marks the point where
the heat transfer switches from electronic conduction in interior WD
layers to photon diffusion in the exterior.

Most WD models rely on semi-analytical descriptions in the
chemical picture~\cite{chab05} where one treats helium as a collection
of stable atoms, ions, and free electrons interacting via approximate
pair potentials. While such approaches work well at low density, they
cannot describe the complex interactions in a very dense system, and a
more fundamental description is required instead. First-principle
methods, such as density functional theory and diffusion quantum Monte
Carlo, are necessary as they provide a full accounting of quantum and
statistical laws that govern the electrons and nuclei.

\begin{figure}[!t]
\includegraphics[width=.5\textwidth]{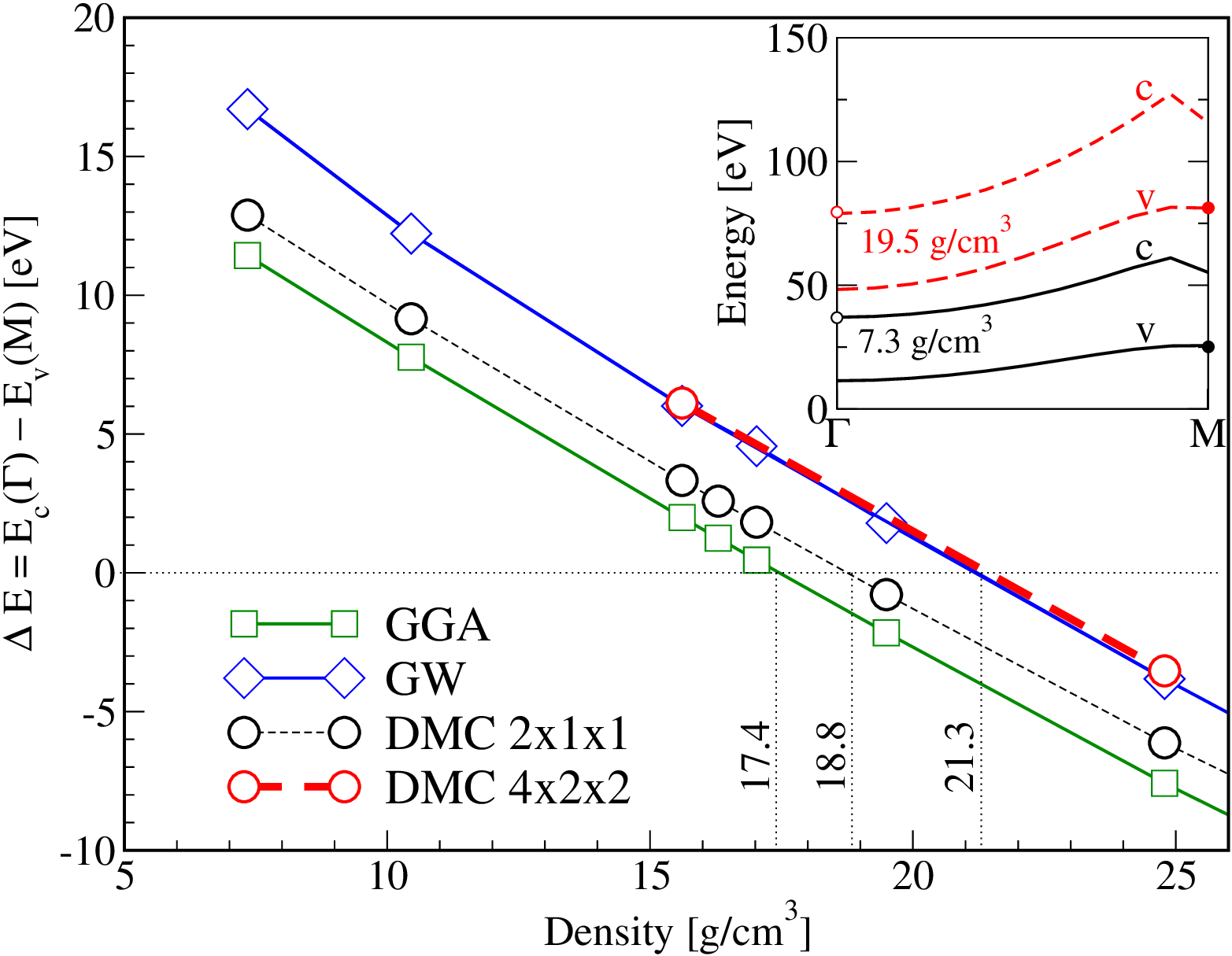}
\caption{\label{Egap}Comparison of band gaps as function of density. DMC results for
  a small 2x1x1 rectangular supercell lie parallel and above the
  GGA gaps by $1.4$ eV. DMC (4x2x2 rectangular cell) and GW
  metallization density of $21.3(1)$ g/cm$^3$ are in
  agreement. GGA underestimates the band gap by about $4$ eV. The
  inset shows a part of the electronic band structure including the
  indirect gap. The filled and open circles indicate, respectively,
  the highest occupied valence state at the M point and the lowest
  unoccupied conduction state at $\Gamma$. Two densities are
  shown. The insulating state (black curve) lies below the metallic
  state (red dashed curve). }
\end{figure}

Recent calculations by Kietzmann \textit{et al.}~\cite{redmer} relied
on DFT and the generalized gradient approximation (GGA)~\cite{gga} to
calculate the electrical conductivity and locate the
insulator-to-metal transition in dense fluid helium. Kowalski
\textit{et al.}~\cite{Kowalski} went beyond GGA to study the EOS, the
electrical and optical properties of fluid helium up to $2$
g/cm$^{3}$. Stixrude and Jeanloz~\cite{loz08} studied the band gap
closure in fluid helium over a wide range of densities including
conditions of giant planet interiors.

In this Letter, we study the metallization in solid helium at
densities above $2$ g/cm$^3$ using several first-principle simulation
techniques. Our intention is to give an assessment of the accuracy of
the widely used GGA for calculating total energies and band gaps
at extreme conditions. For this reason, we use the accurate, but
expensive diffusion quantum Monte Carlo (DMC) method and compare with
the GW approximation to correct the GGA band gaps. We also derive
the equation of state (EOS) and study the effect of the zero point
motion of the nuclei on the band gap closure. We expect our
first-principle study to serve as a guide for future laser
experiments that are planned to extend the EOS measurements to very high
pressures~\cite{nellisloz}.

Our DFT calculations were performed with the \textit{ABINIT}
plane-wave basis code~\cite{abinit}. The electron-nuclei interactions
were treated by a local Troullier-Martin norm-conserving
pseudopotential~\cite{fhi98pp} with a core radius of 0.4~a.u. We use
Perdew-Burke-Ernzerhof GGA~\cite{gga} for the exchange-correlation
functional. We worked with an 8x8x8 Monkhorst-Pack k-point grid and a
plane wave energy cutoff of $230$ Ha.

Mao \textit{et al.}~\cite{mao93} have demonstrated experimentally that
solid helium still remains in the hcp structure up to high pressures
($57$ GPa), apart from a limited fcc loop along the melting line at
low temperatures. So we adopt the hcp solid phase and optimize the
cell geometry iteratively until the pressure has converged to within
$1\%$.

The band structure in the inset in Fig.~\ref{Egap} shows that solid
helium in the hcp structure exhibits an indirect band gap. The lowest
unoccupied state occurs at the $\Gamma$ wave vector. The highest
occupied state is at $\mathbf{k}=0.95\mathbf{k}_{\rm M}$, which is very close
to the M wave vector. Subsequently, we approximate the excitation gap
by the difference in energy between the valence M point and the
conduction $\Gamma$ point. As Fig.~\ref{Egap} shows, the band gap
decreases almost linearly with density. GGA predicts the band gap
closure at a density of $17.4$ g/cm$^3$, which corresponds to a
pressure of $17.0$ TPa.

Kohn-Sham GGA is known to systematically underestimate the band
gap. 
DMC is expected to predict the band gap width more accurately, because
it explicitly includes electronic correlation effects. In fact, DMC
has been used successfully to describe the electronic ground state in
weakly as well as in strongly correlated
systems~\cite{foulkes}. Excited states were also calculated reliably
with DMC~\cite{Martin94,needs06}. 
One needs a large supercell to describe all correlation effects, which
comes at a high computational cost since DMC scales as high as
$O(aN^3+bN^2)$, where $N$ is the number of electrons.

Our DMC simulations were performed with the \textit{CASINO}
code~\cite{casino}. In DMC, the Schr\"{o}dinger equation is solved
stochastically by simulating branching and diffusion in imaginary
time. A trial wave function enters in the propagation of electronic
configurations and we use the fixed-node approximation to
avoid the fermion sign problem.  Our trial wave function is of the
Slater-Jastrow form. The parameters in the Jastrow factor were
optimized by variance minimization. They comprise electron-electron,
electron-nucleus and electron-electron-nucleus terms. 
that the
We use GGA orbitals for the Slater part. We also keep the same
pseudopotential as in DFT calculations. We picked a conservative high
energy cutoff of $800$ Ha. For efficiency, the orbitals are
represented numerically using blip functions~\cite{alf04}. Our results
are well converged with a time step of $0.002$ a.u. To calculate the
band gap, we promote one electron, either spin-up or -down, from the
valence band to the conduction band ($M \longrightarrow \Gamma$). The
calculations for the excited state used the same Jastrow parameters as
for the ground state calculation because the DMC energy depends only
on the nodes of the trial wavefunction that are determined by the
Slater determinant.  To include the $\Gamma$ and M wave vectors into
one DMC band gap calculation, we have chosen rectangular 2x1x1 or
4x2x2 supercells with, respectively 16 or 128 electrons. We also
performed simulations with up to 3x3x3 triangular supercells with 108
electrons to determine the ground state energies and to study their
finite-size dependence.

Figure~\ref{Egap} shows the DMC band gaps to be larger than the GGA
results, as expected. The DMC curves for the 2x1x1 and 4x2x2
rectangular supercells lie parallel to the GGA curve, hence showing a gap
correction that is independent of density. DMC simulations in a 4x2x2
rectangular cell predict a metallization density of $21.3(1)$
g/cm$^3$.  We consider the 4x2x2 gap results to be converged with
respect to system size because the ground state energy agrees well with
triangular 3x3x3 supercells and the remaining finite size corrections
are small.

The 4x2x2 DMC band gaps agree very well with the GW results, which
also show a linear behavior with density similar to all previous
calculations. The GW approximation has proven to be a reliable method
for correcting the GGA band gaps in a variety of
materials~\cite{johnson}. Our GW band gap corrections are calculated
within an accuracy of $0.1$ eV after converging the number of bands
($50$) and plane waves ($27$ Ha). In comparison, our metallization
density is significantly higher than the linear-muffin-tin-orbitals
prediction of $13.5$ g/cm$^3$ for helium in the fcc
phase~\cite{marvin}.

We tried improving the nodes of the trial wave function in DMC
by adding a backflow correction to the Slater determinant. This method
introduces further correlation to the trial wave function by replacing the
electron coordinates in the determinant by a set of collective
coordinates. The DMC total energy decreased slightly but the band
gap did not change within error bars ($0.02$ eV). 

We also studied whether the zero point motion of the nuclei has any
effect on the metallization density because one might expect the
disorder introduced by the zero point motion to reduce the
metallization density, as already noted in the
fluid~\cite{Kowalski,loz08}. We generated series of configurations
with path integral Monte Carlo (PIMC)~\cite{pimc} simulations for
different temperatures between 500 and 5000~K. The helium atoms
interact with an effective pair potential that we constructed by
matching the forces~\cite{forcematching} of a density functional
molecular dynamics (DFT-MD) simulation at 5000~K. We verified the
accuracy of the potential by comparing the original DFT-MD pair
correlation function with that of classical Monte Carlo
simulations. We then computed the GGA band gaps for the PIMC
configurations and compare with perfect hcp lattice results. Within
error bars, the zero point motion had no effect on the band gap.

\begin{figure}[!t]
\includegraphics[width=.4\textwidth]{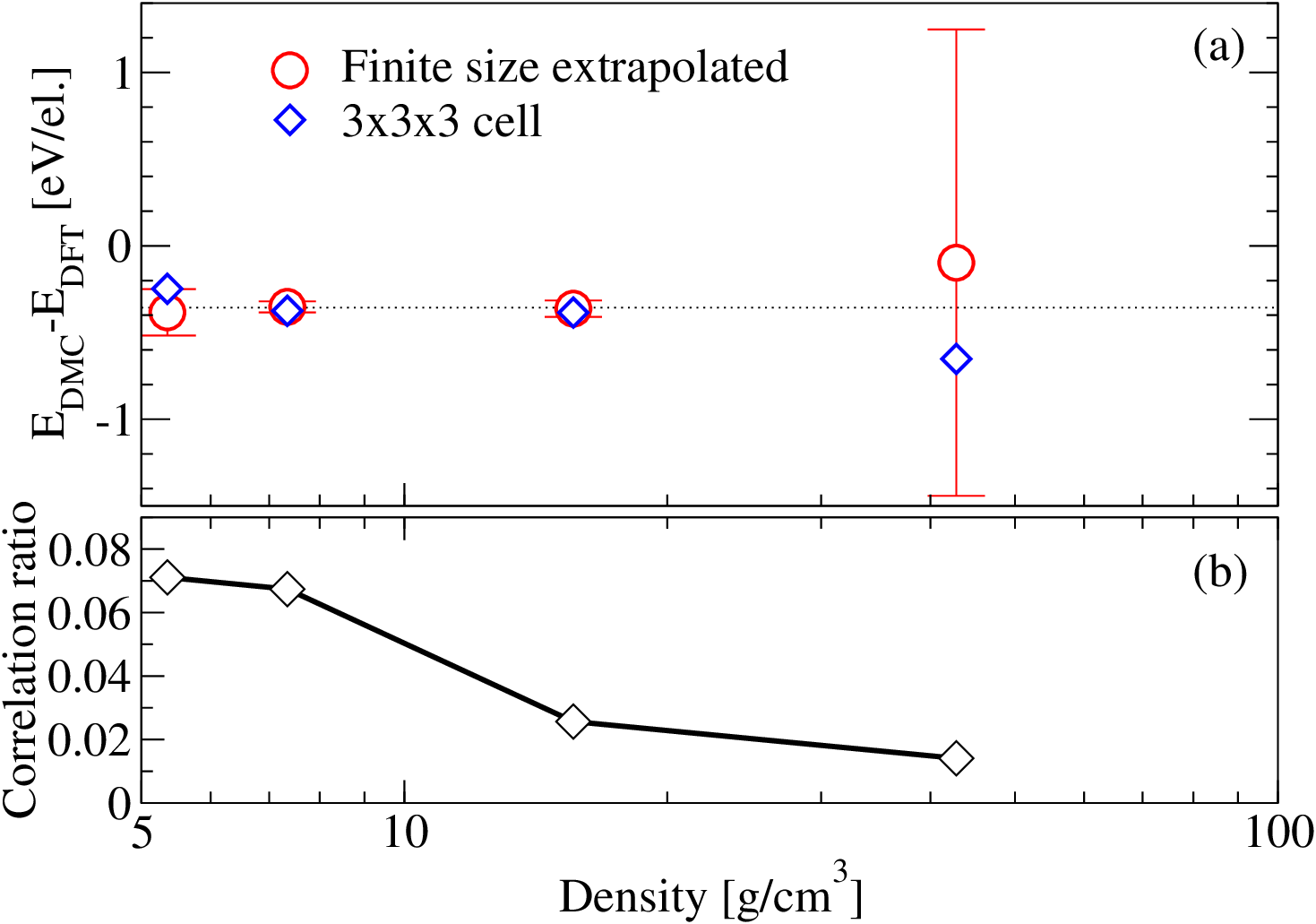}
\caption{\label{dftqmcer} Panel (a): Difference between DMC and GGA 
  ground state energies. We calculate the DMC energy in two ways, by
  extrapolations to infinite cell size (circles) and by directly using
  results from our largest 3x3x3 triangular supercell (diamonds). In
  panel (b), we relate the DFT-DMC energy difference to the amount of
  mechanical work needed to reach a certain density. We plot the
  correlation fraction, $(E_{\rm DFT}-E_{\rm DMC})/(E_{\rm DFT} -
  E_{\rm atom})$, where $E_{\rm atom}=-79.0048$~eV is the exact energy
  of the isolate helium atom.}
\end{figure}

DMC is computationally intensive which limits the size of the
supercell used to simulate the infinite solid. Our biggest supercells
consist of 3x3x3 triangular ($108$ electrons) and 4x2x2 rectangular
($128$ electrons) primitive unit cells. We correct the energies by
considering the following finite size effects. To first order, the
independent-particle finite size effects (IPFSE)
dominate~\cite{needs99}. The error arises from an incomplete k-point
sampling of the Brillouin zone in the DMC supercell. In DFT, the
computational cost is directly proportional to the number of k-points
because the electrons are treated as independent particles. In DMC,
however, bigger supercells are needed to include more k-points. So the
limitation is more severe since the computational cost in DMC scales
with the number of particles in the supercell as O(aN$^3$ + bN$^2$).
Our IPFSEs range from $0.07~$eV/el up to $2.0~$eV/el with increasing
density in our 3x3x3 triangular cell. We also include kinetic energy
corrections of the order of $0.1~$eV/el, which are due to long range
correlations~\cite{chiesa}.

We correct for Coulomb finite size effects (CFSE) that are due to the
long range interaction between charged particles and their periodic
images. These effects decay with system size as $1/N$ and tend to
lower the energy slightly. We reduce these effects by using the model
interaction potential (MPC) instead of the Ewald
interaction~\cite{foulkes}. The difference between the total ground
state energies computed with these two interactions is $0.04$ eV/el
for the largest volume. The error increases with density because the
periodic image charges are closer in smaller supercells. We obtain
very similar band gaps with the Ewald and the MPC interactions. This
is due to the cancellation of CFSE errors when taking energy
differences to calculate the band gap.

After correcting the DMC ground state energies for finite size
effects, we compare with DFT calculations in Fig.~\ref{dftqmcer}. We
use two ways to estimate the correction energy that is missing in
GGA. First we report the DFT-DMC difference directly for our
larger 3x3x3 supercell. Secondly we extrapolate the DMC energies to
infinite size as function of $1/N$. We derive an uncertainty of the
resulting DMC energies by comparing linear and quadratic
extrapolation. The resulting error bars in Fig.~\ref{dftqmcer} are
small except for the highest density. In the density range of
consideration, the correlation energy error in GGA is approximately
constant, 0.36 eV/el.

The correlation error in GGA becomes less important with increasing
density because it stays constant while the energy increases with
compression. This trend is illustrated in Fig.~\ref{dftqmcer} where we
relate the DFT-DMC energy difference to the energy needed to compress
helium to density, $\rho$. The plotted ratio, $(E_{\rm DFT}-E_{\rm
DMC})/(E_{\rm DFT} - E_{\rm atom})$, approaches unity in the low
density limit. In the opposite high density limit, the graph shows
how it tends to zero as helium approaches the state of a
one-component plasma with a neutralizing background. The kinetic
energy of homogeneous electron gas dominates the correlation and
eventually the Coulombic energy terms. GGA is expected to describe
this limit well since the exchange-correlation functional was derived from
DMC simulations of the homogeneous electron gas~\cite{exc}.

\begin{figure}[!t]
\includegraphics[width=.45\textwidth]{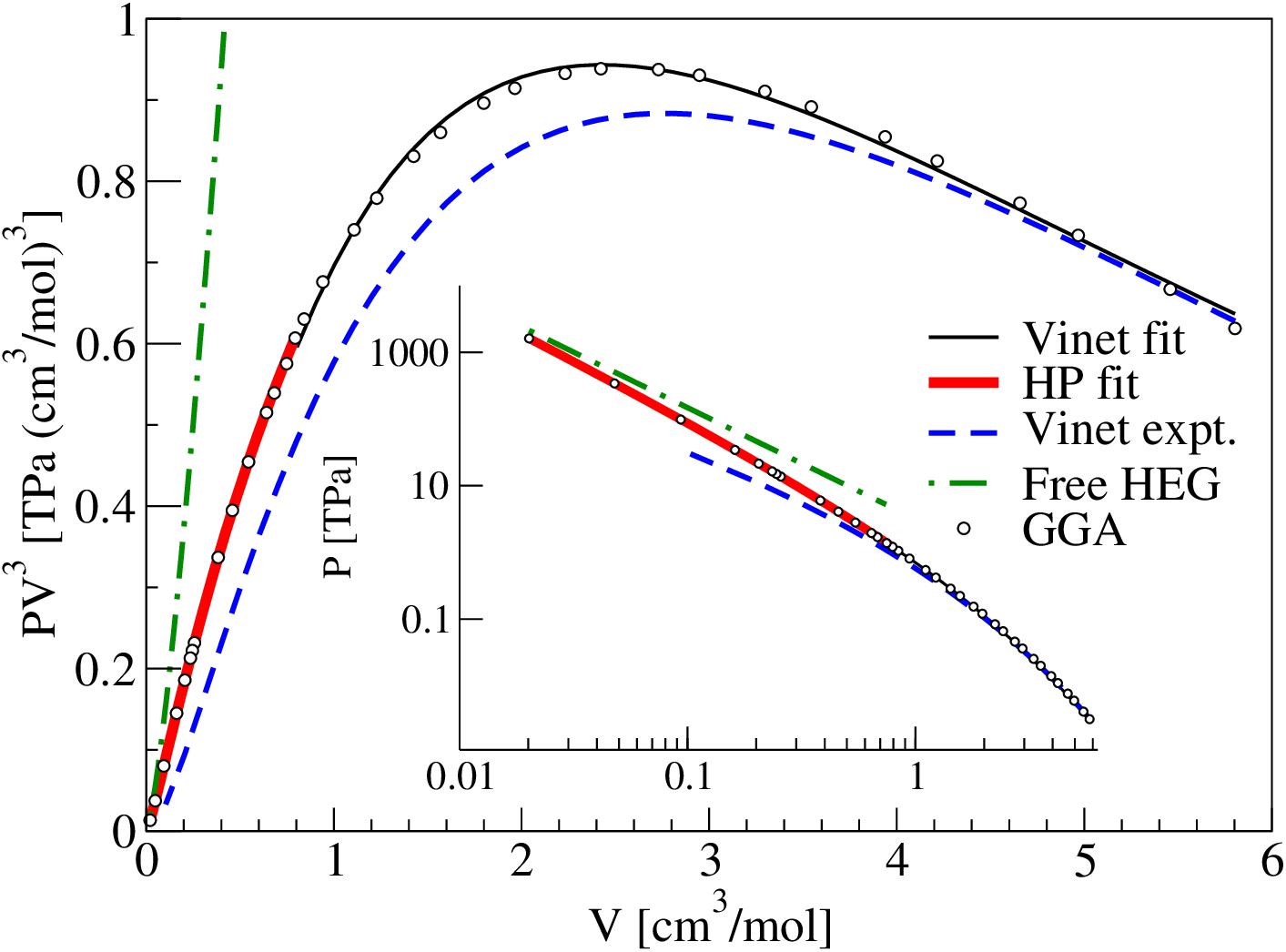}
\caption{\label{eos} Pressure-volume relationship derived from 
       GGA calculations for solid helium. The experimental Vinet
       fit~\cite{loubeyre}, valid up to 57 GPa, agrees with the low
       pressure Vinet fit of GGA data. The high pressure (HP) fit to
       the GGA data approaches the free homogeneous electron gas (HEG). To emphasize
       differences, $PV^3$ was plotted in the main graph.}
\end{figure}

Since the corrections to the GGA energy appear to be independent of
density, there are no corrections to pressures derived from GGA. We
were able to represent our zero-temperature static lattice GGA
pressure data in the insulating regime by a Vinet EOS
curve~\cite{Vinet} with the parameters $V_0=20.6397$ cm$^3$/mol as
zero-pressure volume, $B_0=0.01928$ GPa as bulk modulus, and
$B'_0=9.2153$ as its derivative. The fit reproduced our GGA data from
3 to 1200 GPa with an accuracy of 3\%. The comparison with
experiments~\cite{loubeyre} was studied in great detail
earlier~\cite{lowPsim}.

It was not possible to extend the Vinet fit into the metallic
regime. Instead, we adopt a fit based on the
parametrization of the homogeneous electron gas energy. This fit
includes the kinetic, Coulombic, exchange as well as correlation terms
and, in this case also ionic contributions, $P(V) =
\frac{a_1}{V^{5/3}}+\frac{a_2}{V^{4/3}} + \frac{a_3}{V}+
\frac{a_4}{V^{2/3}}$. In units of GPa and cm$^3$/mol, the coefficients are 
$a_1=3186.21$, $a_2=-2761.74$, $a_3= -565.78 $, and $a_4=854.71$ where the
leading coefficient is taken from the free Fermi gas. The fit
reproduces our DFT data points from 1.2 to 1600 TPa within
$0.5\%$. In Fig.~\ref{eos}, we show the pressure over a large density
range. In the high density limit, the correlation effects decrease and
the DFT pressure approaches the free homogeneous electron gas
behavior.

In conclusion, we have demonstrated that solid helium reaches a
metallic state at an extreme pressure of $25.7$ TPa, which is
significantly larger than predicted by standard GGA method. For WD
interiors, this implies that the inner layer of metallic helium is
thinner and the outer region where photon diffusion dominates the heat
transport is larger than previously predicted. 

With quantum Monte Carlo we have shown that standard GGA methods
underestimate the band gap in solid helium by $4$ eV, which translates
into an underestimation of the metallization pressure by $40\%$. The
GW band gap corrections are in good agreement with DMC calculations,
which offers the possibility of using GW for correcting the band gaps
derived from GGA simulation of fluid helium at high pressure and to
make more realistic comparisons with shock wave measurements of
conductivity and reflectivity.

Finally, we determined the equation of state and presented a fit. We
analyzed the correlation effects that are missing in GGA and
demonstrated that their importance decreases relative to the total
energy with increasing density as helium approaches the state of a
one-component plasma with a rigid neutralizing background.

This work was supported by NSF and NASA. We thank the Cambridge group
for the CASINO code. NERSC provided computational resources. We
acknowlegde useful discussions with E. Quataert, P. Chang, and
R. Jeanloz.


\end{document}